\documentclass[final,5p,times,twocolumn]{elsarticle}
\usepackage{graphicx}
\usepackage{amssymb}
\usepackage{amsmath}
\usepackage{amsthm}
\usepackage{lineno}
\usepackage{bm}% bold math
\newcommand{\beq}{\begin{equation}}
\newcommand{\eeq}{\end{equation}}
\newcommand{\beqn}{\begin{equation*}}
\newcommand{\eeqn}{\end{equation*}}
\newcommand{\ba}{\begin{array}}
\newcommand{\ea}{\end{array}}
\newcommand\gr[1]{\mathrm{#1}}    % font for groups etc.
\newcommand\muI{\mu_{\mathrm{I}}}
\newcommand{\ie}{{i.e.}}

\DeclareRobustCommand\alt{\lesssim}
\journal{Physics Letters B}

\begin{document}
\begin{frontmatter}

\title{Ginzburg-Landau phase diagram of QCD near chiral critical point\\
-- chiral defect lattice and solitonic pion condensate --}
\author{Hiroaki Abuki}
\address{Department of Physics, Tokyo University of Science, Kagurazaka
 1-3, Shinjuku, Tokyo 162-0825}

\begin{abstract}
We investigate the influence of the isospin asymmetry on the phase
 structure of quark matter near the chiral critical point systematically
 using a generalized version of Ginzburg-Landau approach.
The effect has proven to be so profound that it brings about not only a
 shift of the critical point but also a rich variety of phases in its
 neighborhood. 
In particular, there shows up a phase with spatially varying charged
 pion condensate which we name the ``solitonic pion condensate'' in
 addition to the ``chiral defect lattice'' where the chiral condensate
 is partially destructed by periodic placements of two-dimensional
 wall-like defects.
Our results suggest that there may be an island of solitonic pion
 condensate in the low temperature and high density side of QCD phase
 diagram.
\end{abstract}

\begin{keyword}
Chiral Symmetry breaking \sep Charged pion condensate \sep Inhomogeneous
 condensates

\end{keyword}

\end{frontmatter}

%\linenumbers

\emph{Introduction.--}
The chiral critical point (CP) in QCD phase diagram is the subject of
extensive theoretical/experimental studies \cite{Fukushima:2010bq}. 
It was shown in \cite{Nickel:2009ke,Nickel:2009wj} that once the
possibility of inhomogeneity is taken into account, the CP turns
into a Lifshitz critical point (LCP) where 
a line of the chiral crossover meets two lines of second-order phase
transitions surrounding the phase of an inhomogeneous chiral
condensate.
The inhomogeneous state can be viewed as an ordered phase separation,
produced via the compromise between quark-antiquark
attraction and a pair breaking due to imbalanced population of quarks
to antiquarks \cite{Abuki:2011pf,Fukushima:2012mz}.
Such inhomogeneity appears rather commonly in a wide range of physics;
the Abrikosov lattice \cite{Abrikosov} and the
Fulde-Ferrell-Larkin-Ovchinnikov superconductors \cite{FFLO}
are such examples.

In this Letter, we address the question what is the possible impact of
the effect of an isospin asymmetry on the LCP. 
For bulk systems such as matter realized in compact stars, 
such flavor symmetry breaking is caused mainly by a neutrality
constraint that should be
imposed to prevent the diverging energy density.
The effect leads to a rich variety of color superconducting
phases at high density \cite{Abuki:2004zk}. 
On the other hand, at large isospin density
QCD vacuum develops a charged pion condensate (PC) as soon as
$|\muI|>m_\pi$ with $m_\pi$ and $\muI$ being the vacuum pion mass and
the isospin chemical potential \cite{Son:2000xc}.
The PC has a rich physical content including a crossover from a
Bose-Einstein condensate of pions to a superfluidity of the
Bardeen-Cooper-Schrieffer type, and has been extensively studied
using effective models \cite{He:2005nk}.

We focus here how the neighborhood of CP is to be modified by inclusion
of isospin density. 
To this aim, we use the generalized Ginzburg-Landau (GL)
approach developed in \cite{Nickel:2009ke,Abuki:2011pf} which can
give rather model-independent predictions near the CP.
Since we are interested in the response of the CP and its vicinity
 against $\muI\ne0$, our strategy is to take $\muI$ as a
 perturbative field and expand the GL functional with respect to it.
The inclusion of $\muI$ further brings new GL parameters, but they
can be evaluated within the quark loop approximation
 \cite{Nickel:2009ke,Abuki:2011pf} since gluons are insensitive
 to isospin charge.
What we will find is that the isospin asymmetry dramatically modifies
the neighborhood of CP bringing about new multicritical points.
Accordingly, the inhomogeneous version of charged pion condensate
dominates a major part of phase diagram. 

\emph{Generalized Ginzburg-Landau approach.--}
We consider two-flavor QCD, and assume the existence of a
tricritical point (TCP) located in the $(\mu,T)$-phase diagram
in the chiral limit at vanishing $\muI$.
We take the chiral four vector $\phi=(\sigma,\bm{\pi})\sim%
(\langle\bar{q}q\rangle,\langle\bar{q}i\gamma_5\bm{\tau}q\rangle)$
as a relevant order parameter of the system.
A minimal GL description of TCP requires the expansion of the
thermodynamic potential up to sixth order in $\phi$.
The resulting chiral O(4) invariant potential
expanded up to the sixth order is, with incorporating the derivative
terms as well \cite{Nickel:2009ke,Abuki:2011pf}:
$\omega[\phi(\textbf{x})]=\sum_{n=1,2,3}\omega_{2n}[\phi(\textbf{x})],
$ where
\beq
\begin{array}{l}
 \omega_2[\phi(\textbf{x})]=\frac{\alpha_2}{2}\phi(\textbf{x})^2,\;%
 \omega_4[\phi(\textbf{x})]=\frac{\alpha_4}{4}\left(\phi^4%
 +(\nabla\phi)^2\right),\\[1ex]
\begin{split}
\textstyle
 \omega_6[\phi(\textbf{x})]=\,&\textstyle\frac{\alpha_6}{6}\Big(\phi^6%
+3\big[\phi^2(\nabla\phi)^2-(\phi,\nabla\phi)^2\big]\\
&\textstyle+5(\phi,\nabla\phi)^2%
+\frac{1}{2}(\Delta\phi)^2\Big).
\end{split}
\end{array}
\label{eq:omega}
\eeq
The current quark mass adds to this a term
$\omega_1[\sigma(\textbf{x})]=-h\sigma(\textbf{x})$ which explicitly
breaks O(4) symmetry down to O(3), and thus makes the condensate align
in the direction $\phi\to(\sigma,{\textbf 0})$. 
We use $\alpha_6^{-1/2}$ as a unit of an energy
dimension.
Accordingly we replace $\alpha_6$ with $1$, and every quantity is to be
regarded as a dimensionless.
Then via scaling $\phi\to\phi h^{1/5}$, $\textbf{x}\to \textbf{x}h^{-1/5}$
together with $\alpha_2\to\alpha_2 h^{4/5}$,
$\alpha_4\to\alpha_4h^{2/5}$, we can get rid of $h$ in
$\omega$ apart from a trivial overall scaling factor $h^{6/5}$, \ie,
$\omega\to\omega h^{6/5}$.
Then we set $h=1$, and retain the original letters $\phi$, ${\textbf x}$,
$\alpha_2$, $\alpha_4$ and $\omega$ hereafter, but we should keep in
mind that they should scale as $h^{1/5}$, $h^{-1/5}$, $h^{4/5}$,
$h^{2/5}$, $h^{6/5}$ respectively.

We assume that $\sigma({\bf x})$ is spatially varying
in one direction, $z$ \cite{Nickel:2009ke}.
The Euler-Lagrange equation (EL), $\delta\omega/\delta\phi(z)=0$, becomes
\beq
\begin{array}{rcl}
  6h&=&\sigma^{(4)}(z)%
-10(\sigma^2\sigma^{\prime\prime}+\sigma(\sigma^{\prime})^2)-3\alpha_4\sigma^{\prime\prime}\\[1ex]
 &&+6\sigma^5+6\alpha_4\sigma^3+6\alpha_2\sigma,
\end{array}
\label{eq:EL}
\eeq
where $h$ is temporarily recovered to remind us that the
term comes from the mass term.
We try the ansatz \cite{Nickel:2009wj}
\beq
 \sigma(z)=A\mathrm{sn}(kz-b/2,\nu)\mathrm{sn}(kz+b/2,\nu)+B,
\label{eq:ansatz}
\eeq
where ``$\mathrm{sn}$'' is the Jacobi elliptic function with $\nu$
the elliptic modulus, and $k$, $b$, $A$, $B$ are real parameters.
We call the state the ``chiral defect lattice''
(CDL)\footnote{The ansatz is called the ``solitonic chiral condensate'' in
\cite{Nickel:2009wj}. 
As we will see later, the state can be viewed as a periodically placed
wall-like defects of chiral condensate, so we use the term ``CDL'' here.}.
This is a spatially modulating state having a period $\ell_p=2K(\nu)/k$.
Let us first show that the ansatz actually provides a one-parameter
family of solution to the EL (\ref{eq:EL}) when suitable conditions for $A$,
$B$, $k$ and $b$ are met.
First, we note from (\ref{eq:ansatz}),
$\mathrm{sn}(kz,\nu)^2=\frac{(\sigma(z)-B)/A+b_2}{1+\nu
b_2(\sigma(z)-B)/A}$
with $b_2\equiv\mathrm{sn}(b/2,\nu)$.
$f(z)=\mathrm{sn}(kz,\nu)$ obeys the Jacobi
differential equation $(f^{\prime})^2=k^2(1-f^2)(1-\nu^2 f^2)$, which
translates into 
\beq
 d_0=(\sigma^{\prime})^2+d_1 \sigma+d_2 \sigma^2+d_3 \sigma^3+d_4 \sigma^4,
\label{eq:diff1}
\eeq
where $\{d_0,d_1,d_2,d_3,d_4\}$ are functions of $A$, $B$, $b$, $k$
and $\nu$. 
We here give the expressions for $d_3$ and $d_4$ only,
\beq
 \begin{array}{rcl}
  d_3=4d_4\Big(A\frac{\mathrm{cn}(b,\nu)%
\mathrm{dn}(b,\nu)}{\nu^2\mathrm{sn}^2(b,\nu)}-B\Big),\,%
d_4=-\frac{k^2\nu^4\mathrm{sn}^2(b,\nu)}{A^2}.
 \end{array}
\label{eq:d34}
\eeq
Differentiating (\ref{eq:diff1}) with respect to $z$ and dividing
the result by $2\sigma^\prime$, we obtain
\beq
\textstyle
-\frac{d_1}{2}=\sigma^{\prime\prime}(z)+d_2\sigma(z)+\frac{3d_3}{2}\sigma(z)^2+2d_4\sigma(z)^3.
\label{eq:diff2}
\eeq
Differentiating this twice we have
\beq
\begin{array}{rcl}
 0&=&\sigma^{(4)}(z)+6d_4\sigma^2\sigma^{\prime\prime}+12d_4\sigma(\sigma^\prime)^2%
+d_2\sigma^{\prime\prime}\\[1ex]
 &&+3d_3(\sigma^\prime)^2+3d_3\sigma\sigma^{\prime\prime}.
\end{array}
\label{eq:diff3}
\eeq
Adding to this, $(f_0+f_1\sigma(z))\times$(\ref{eq:diff1}) 
and $(g_0+g_1\sigma(z)+g_2\sigma(z)^2)\times$(\ref{eq:diff2}) with
${f_0,f_1,g_0,g_1,g_2}$ being arbitrary
constants, we obtain a wider fourth-order differential equation.
Then by tuning $f_0=g_1=-3d_3$, we can get rid of unnecessary 
${\sigma^\prime}^2$ and $\sigma\sigma^{\prime\prime}$ terms, and setting
$f_1=-10-12d_4$, $g_2=-10-6d_4$, $g_0=-d_2-3\alpha_4$ leads to
\beq
\begin{split}
\gamma(\{d_{i}\},\alpha_4)=\,&\textstyle
 \sigma^{(4)}(z)-10(\sigma^2\sigma^{\prime\prime}+\sigma(\sigma^\prime)^2)-3\alpha_4\sigma^{\prime\prime}\\
&\textstyle-6d_4(5+4d_4)\sigma^5-5d_3(5+6d_4)\sigma^4\\
&\textstyle+\sum_{n=3,2,1}\beta_n(\{d_i\},\alpha_4)\sigma^n,
\end{split}
\label{eq:diff4}
\eeq
where $\gamma$ and $\beta_n$ ($n=1,2,3$) are simple algebraic functions
of $d_0,d_1,d_2,d_3,d_4$, and $\alpha_4$.
Matching the coefficients of $\sigma^5$ and $\sigma^4$ with those in
(\ref{eq:EL}) leaves two choices; $(d_3,d_4)=(0,-1)$ or $(0,-1/4)$. 
It turns out that the latter cannot satisfy the remaining constraints 
so we choose $(d_3,d_4)=(0,-1)$ which, with (\ref{eq:d34}), constrains
$A$ and $B$ as
\beq
\textstyle
A=k\nu^2\mathrm{sn}(b,\nu),\,B=k\frac{\mathrm{cn}(b,\nu)\mathrm{dn}(b,\nu)}%
{\mathrm{sn}(b,\nu)}.
\label{eq:AB}
\eeq
The conditions $\beta_3=6\alpha_2$ and $\beta_2=0$ are 
then automatically satisfied, so we are left with two constraints
$6h=\gamma%\equiv\frac{d_1(3\alpha_4+d_2)}{2}
$ 
and $6\alpha_2=\beta_1%\equiv 2d_0-d_2(3\alpha_4+d_2)
$.
Now that $\{d_i\}$ are functions of three variables $\{k,\nu,b\}$, the two
conditions fix two of them, for instance, 
$\{k,b\}$ at a fixed elliptic modulus $\nu$. 
Hence, the ansatz (\ref{eq:ansatz}) together with (\ref{eq:AB}) gives a
one-parameter solution to (\ref{eq:EL}).
To our knowledge, this is the first demonstration of the fact that
(\ref{eq:ansatz}) constitutes a solution also in the GL functional
approach which could be applied in a wide range of physics.
The parameter $\nu$ is to be determined via the minimization of
thermodynamic potential $\Omega$, the spatial average 
of the energy density over one period $\ell_p=2K(\nu)/k$:
\beq
\textstyle
 \Omega(\nu;\alpha_2,\alpha_4)%
 =\frac{1}{\ell_p}\int_{-\ell_p/2}^{\ell_p/2}dz\,\omega[\sigma(z)].
\label{eq:wage}
\eeq

Let us briefly check the two extreme limits, $\nu\to1$ and $\nu\to0$.
First when $\nu\to1$,
\beq
\textstyle
 \sigma(z)\to \sigma_{\gr{sd}}(z)=\frac{k}{\mathrm{th}(b)}\left(1%
-\mathrm{th}^2(b)f_{\mathrm{def.}}(kz,b)\right),
\label{eq:singledefect}
\eeq
where the subscript ``sd'' refers to a ``single-defect'', and
$f_{\mathrm{def.}}(kz,b)\equiv 1-\gr{th}(kz+b/2)\gr{th}(kz-b/2)$.
This describes a defect in chiral condensate, represented by a
\emph{soliton-antisoliton pair} located at $z=0$.
The homogeneous value gets eventually recovered as $|z|\to\infty$:
$\sigma_\gr{sd}(\pm\infty)\equiv\sigma_L=k/\gr{th}(b)$.
Since $k=\sigma_L\gr{th}(b)$, we regard $\sigma_{\gr{sd}}(z)$ as a
function of $z$ parametrized by $\sigma_L$ and $b$. On the other hand,
when $\nu\to 0$ the ansatz reduces to, retaining up to the first
non-trivial order in $\nu$,
\beq
\textstyle
\sigma(z)\to
\sigma_{\gr{sin}}(z)=k\cot(b)-\nu^2\frac{k\sin(b)}{2}\cos(2kz).
\label{eq:ripping}
\eeq
This is the state where chiral condensate is about to develop a ripple
sinusoidal wave on the homogeneous background. We denote the background
chiral condensate as $k\cot(b)\equiv \sigma_S$, $\sigma_{\gr{sin}}(z)$
is now viewed as a function of $z$ parametrized by $\sigma_S$, $k$, and
vanishing $\nu$.
%---------------------------FIGURE---------------------------%
\begin{figure}[t]
\centerline{
\includegraphics[width=0.35\textwidth]{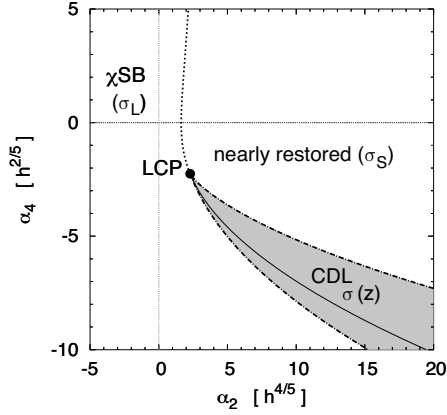}
}
\caption{The GL phase diagram at $\muI=0$. 
The CDL is realized in the shaded region.
The dotted line starting from the LCP
 $(\alpha_2^{\gr{LCP}},\alpha_4^{\gr{LCP}})%
\equiv\left(\frac{5}{4}\frac{3^{4/5}}{2^{2/5}},%
-\frac{5}{2^{1/5}3^{3/5}}\right)$  
stands for the crossover,
while two dot-dashed lines enclosing the CDL
denote the second-order phase transitions.
}
\label{fig:pd0}
\end{figure}
%---------------------------FIGURE---------------------------%

\emph{Phase structure at $\muI=0$.--}
We compute the phase diagram via minimization of (\ref{eq:wage}).
The result is displayed in Fig.~\ref{fig:pd0}.
The CP is indeed realized as the LCP 
where the three phases meet; the CDL phase with $\sigma(z)$,
chiral symmetry broken $(\chi\gr{SB})$ phase with an homogeneous 
condensate $\sigma_{L}$,
and the nearly symmetry-restored phase characterized by a smaller condensate
$\sigma_{S}$.
For illustration, also shown by a solid line is the line of would-be
first-order transition.
%---------------------------FIGURE---------------------------%
\begin{figure}[tp]
\centerline{
\includegraphics[width=0.48\textwidth]{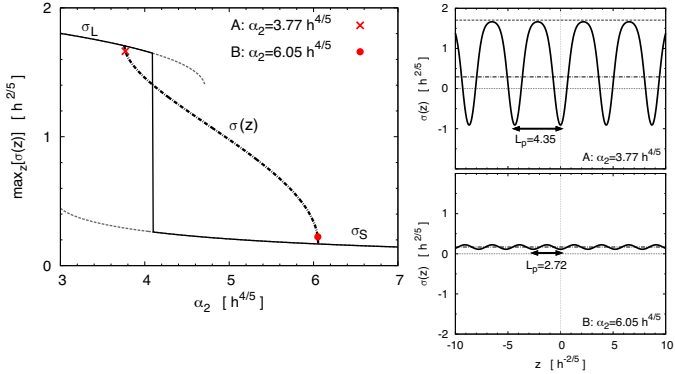}
}
\caption{
(Left panel):~The amplitude $\gr{max}_z[\sigma(z)]$ as a function of
 $\alpha_2$ along the line $\alpha_4=-4$.
(Right panel):~The spatial profiles of $\sigma(z)$ at point {\bf A}
 ($\alpha_2=3.77h^{4/5}$)
and {\bf B} ($\alpha_2=6.05h^{4/5}$) shown in the left figure by a cross
 and a circle.
}
\label{fig:pr1}
\end{figure}
%---------------------------FIGURE---------------------------%
Fig.~\ref{fig:pr1} shows how $\sigma(z)$ smoothly interpolates between
$\sigma_L$ and $\sigma_S$ along $\alpha_4=-4$.
Displayed in the left panel is the max amplitude $\max_z[\sigma(z)]$ 
as a function of $\alpha_2$.
Abrupt drop in $\sigma$ indicated by a solid line
shows the location of would-be first-order transition which would
have taken place if we ignored the possibility of inhomogeneity.
Two figures in the right panel show the spatial profiles of $\sigma(z)$
for two values of $\alpha_2$ denoted by {\bf A} and {\bf B}, whose
locations are marked by a cross and a circle in the left figure.
It looks like the periodic placements of defects 
near the $\sigma_L$-side, while it is just a tiny ripple sinusoidal wave
near the $\sigma_S$-side.
For both points, the $\sigma_S$-state exists 
as a local minimum, and its magnitude is depicted by a
dot-dashed line.
We see $\sigma_S$ is roughly the median of $\sigma(z)$.
At point {\bf A}, the $\sigma_L$-state also exists as a metastable
state. The magnitude of $\sigma_L$ is shown by a dashed line, which
roughly corresponds to the max amplitude of $\sigma(z)$. 
The modulation period, $\ell_p$, is also shown by the arrow. 
$\ell_p$ grows towards the $\sigma_L$-side, evolving
to infinity realizing a single defect state
$\sigma_{\gr{sd}}(z)$.

\emph{Formation of a single defect.--}
The second-order phase transition from the $\chi\gr{SB}$ phase to
the CDL is signaled by the formation of a single
wall-like defect (\ref{eq:singledefect}), that is, a creation of
soliton-antisoliton pair.
Let us briefly describe this critical condition.
When the defect forms in the sea of
homogeneous background, mass per unit area associated with 
the wall extending in the transverse $(x,y)$-plane should vanish. 
The energy per unit area is
$f_{\gr{sd}}(b;\alpha_2,\alpha_4)%
=\int_{-\infty}^{\infty}dz\left(\omega[\sigma_{\gr{sd}}(z)]%
-\omega[\sigma_L]\right)$.
Note that, for any $(\alpha_2,\alpha_4)$, once the
homogeneous value $\sigma_L$ is numerically fixed, 
$f_{\gr{sd}}$ is a function of $b$ only.
In Fig.~\ref{fig:pr1}(a), plotted are $f_{\gr{sd}}$ for
$\alpha_2=3.5$, $3.76$, $4.0$ at $\alpha_4=-4$.
We see that $\sigma_{\gr{sd}}(z)$ with $b\ne0$ becomes more
favorable once $\alpha_2$ exceeds $3.76(\equiv\alpha_{2\gr{c}})$, the
critical value for defect formation onset. 
Note that the state with $b=0$ is equivalent to the $\chi\gr{SB}$ as
$\sigma_{\gr{sd}}(z)\to\sigma_L$ with $b\to0$ as seen from
(\ref{eq:singledefect}).
In Fig.~\ref{fig:pr1}(b), the spatial profile of $\sigma_{\gr{sd}}(z)$
at $\alpha_2=\alpha_{2\gr{c}}$ is depicted by a light solid line, and
that of energy density $\omega[\sigma_{\gr{sd}}(z)]-\omega[\sigma_L]$ is
drawn by a heavy solid line.
Contribution from each $\omega_i$ is also separately shown.
We see that the gradient terms in particular in $\omega_6$,
and $\omega_2$ are responsible for the spontaneous generation a defect.

\emph{Rippling of the chiral condensate.--}
To derive the critical line separating the CDL and $\sigma_S$- phases,
we first plug (\ref{eq:ripping}) into (\ref{eq:wage}), then
minimize the result over $b$ and perform expansion about
$\nu^2$. Then looking at the location where the coefficient of $\nu^4$
changes its sign, we reach the condition for the onset of rippling
chiral condensate.

%---------------------------FIGURE---------------------------%
\begin{figure}[t]
\centerline{
\includegraphics[width=0.48\textwidth]{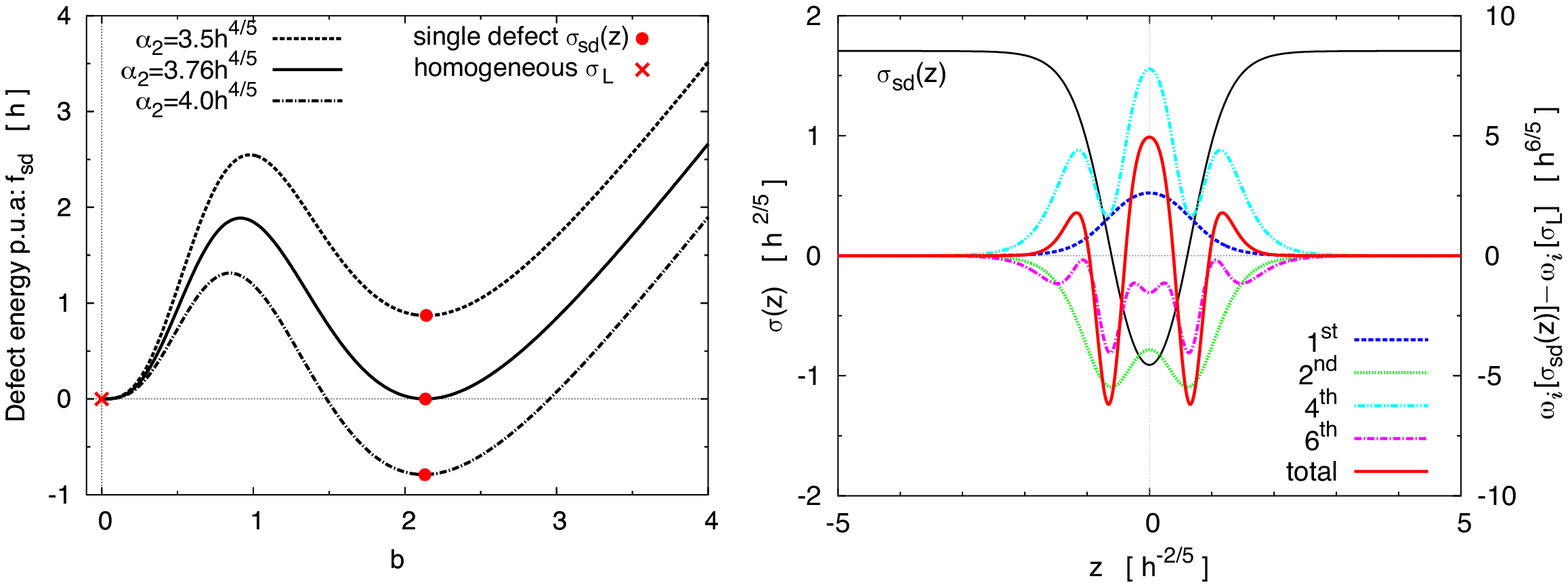}
}
\caption{(Left):~The energy per unit area associated with a
 single wall-like defect as a function of $b$.
(Right):~The spatial profile of $\sigma_{\gr{sd}}(z)$ and
 energy density at the onset $\alpha_2=3.76h^{4/5}$.
}
\label{fig:profile2}
\end{figure}
%---------------------------FIGURE---------------------------%

\emph{Phase structure for $\muI\ne0$.--}
When we take $\muI$ into consideration, the GL coefficients
$\{\alpha_{i}\}$ become functions of $\muI$, 
to be denoted by $\{\alpha_{i}(\muI)\}$. 
In addition, since $\muI$ breaks the isospin
$\gr{SU}(2)_{\gr{V}}$ symmetry to $\gr{U}(1)_{\gr{I}_3}$ which
describes the rotation about the isospin third axis, 
the potential has new feedback terms which is invariant under
$\gr{U}(1)_{\gr{I}_3}$ but not under full $\gr{SU}(2)_{\gr{V}}$. 
Up to the fourth order in $(\sigma,\bm{\pi})$, the most general
form of the feedback potential describing the response to $\muI$ is
\cite{Abuki:2013vwa}
\beqn
\textstyle
\delta\omega_\gr{I}%
=\frac{\beta_2}{2}\bm{\pi}_{\perp}^2%
+\frac{\beta_4}{4}\bm{\pi}_\perp^4%
+\frac{\beta_{4b}}{4}(\phi^2-\bm{\pi}_\perp^2)%
\bm{\pi}_\perp^2+\frac{\beta_{4c}}{4}(\nabla\bm{\pi}_\perp)^2,
\eeqn
where $\bm{\pi}_\perp=(\pi_1,\pi_2)$ is the charged pion doublet.
When $|\bm{\pi}_\perp|\ne0$, the residual
$\gr{U}(1)_{\gr{I}_3}$ (or equivalently the electromagnetic
$\gr{U}(1)_{\gr{Q}}$) gets broken spontaneously.

In order to find an expression of the potential up to
the sixth order in $\{\muI,\sigma,\pi,\nabla\}$ we need to expand
$\alpha_{2n}$, $\beta_{2}$ and $\beta_{4,4b,4c}$ up to the corresponding
orders in $\muI$.
Via explicit computations \cite{Abuki:2013vwa}, we 
have
$\alpha_6(\muI)=\alpha_6+{\mathcal O}(\muI^2)$,
$\alpha_4(\muI)=\alpha_4+\muI^2\alpha_6+{\mathcal O}(\muI^4)$ and
$\alpha_2(\muI)=\alpha_2+{\mathcal O}(\muI^6)$. 
$\beta_{2}$ and $\beta_{4,4b,4c}$ have the following
general structure:
\beqn
\textstyle
\begin{pmatrix}
\beta_2(\muI)\\
\beta_{\{4,4b,4c\}}(\muI)
\end{pmatrix}
=\muI^2\begin{pmatrix}
a & e\muI^2\\
0 & \{b,c,d\} \\
\end{pmatrix}
\begin{pmatrix}
\alpha_4\\
\alpha_6
\end{pmatrix}+\begin{pmatrix}
\mathcal{O}(\muI^6)\\
\mathcal{O}(\muI^4)
\end{pmatrix}.
\eeqn
Straightforward (but tedious) work leads to $a=-\frac{1}{2}$, $e=0$, and
$\{b,c,d\}=\{-2,-2,-\frac{4}{3}\}$.
Plugging all these expressions into the potential
$\omega+\delta\omega_\gr{I}\equiv\omega_\gr{t}$ we have 
\beq
\ba{rcl}
 \omega_\gr{t}&=&\omega_6[\phi(\textbf{x})]-h\sigma+\frac{\alpha_2}{2}\sigma^2%
+\left(\frac{\alpha_2}{2}-\frac{\muI^2\alpha_4}{4}\right)\bm{\pi}_\perp^2\\[1ex]
&&+\left(\frac{\alpha_4}{4}+\frac{\muI^2\alpha_6}{4}\right)\sigma^4%
+\frac{\alpha_4}{2}\sigma^2\bm{\pi}_\perp^2%
+\left(\frac{\alpha_4}{4}-\frac{\muI^2\alpha_6}{12}\right)\bm{\pi}_\perp^4\\[1ex]
&&+\left(\frac{\alpha_4}{4}+\frac{\muI^2\alpha_6}{4}\right)(\nabla\sigma)^2%
+\left(\frac{\alpha_4}{4}-\frac{\muI^2\alpha_6}{12}\right)(\nabla\bm{\pi}_\perp)^2.
\ea
\label{eq:omegamuI}
\eeq
We see $h$ favors condensation in the $\sigma$-direction, while $\muI$
prefers $\bm{\pi}_\perp\ne 0$.
Now the potential has parameters $\{\alpha_2,\alpha_4,\alpha_6,h\}$ and
$\muI^2$.
Repeating the same dimensional and scaling discussion as before, we get
rid of $\alpha_6$ and $h$, so the remaining parameters are
$\alpha_2,\alpha_4$ and $\muI^2$ which scale as $h^{4/5}$, $h^{2/5}$ and
$h^{2/5}$.

%---------------------------FIGURE---------------------------%
\begin{figure}[tbp]
\centerline{
\includegraphics[width=0.48\textwidth]{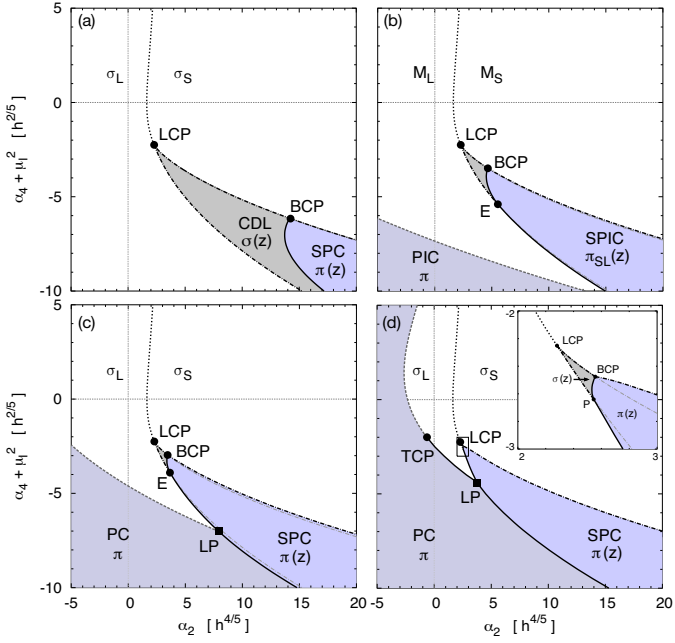}
}
\caption{The GL phase diagram for $\muI^2=0.01$ (a), $0.1$ (b), $0.2$
 (c) and $0.5$ (c). 
The solid lines stand for first-order phase transitions, while
 (dot-)dashed lines represent second-order phase transitions.
}
\label{fig:phase_final}
\end{figure}
%---------------------------FIGURE---------------------------%

The remaining task is to find the most favorable state for a given
parameter set of $\{\alpha_2,\alpha_4,\muI^2\}$.
We here consider four variational states:
\\ \noindent
(i)~The $\chi\gr{SB}$ state with $\sigma\ne0$, $\bm{\pi}_\perp=0$. 
\\ \noindent
(ii)~The CDL state with $\bm{\pi}_\perp=0$, $\sigma(z)$ described in
(\ref{eq:ansatz}). In the same way as before, with a replacement
$\alpha_4\to\alpha_4+\muI^2$ 
we can show this gives a solution to the EL. 
\\ \noindent
(iii)~The homogeneous charged pion condensate (PC) with 
$\bm{\pi}_\perp=(\pi,0)$ and $\sigma\ne0$.
\\ \noindent
(iv)~The solitonic charged pion condensate (SPC) with
\beqn
\textstyle
\sigma\ne0,\;\;\bm{\pi}_\perp=(\pi(z),0),\;\;\pi(z)=k\nu\,\gr{sn}(kz,\nu).
\eeqn
This indeed gives a solution to the EL. 
In Table~\ref{tab:states}, we summarize these states with associated
symmetries \footnote{
We also tried two other exotic inhomogeneous states.
One is the ``skewed chiral spiral'' defined by
$\sigma=\sigma_0+A\sin(kz)$, and $\pi_1=B\cos(kz)$, which is an
extension of the ``CDW'' introduced in \cite{Nakano:2004cd},
and the other is the ``IPC'' state taken in \cite{Gubina:2012wp}.
These were found to be less favorable than the CDL (or SPC) state
for any value of $\muI$.}.

%---------------------------TABLE---------------------------%
\begin{table}[bt]
    \caption{State candidates for $\muI\ne0$.}
\centerline{
\begin{tabular}{|c|c|c|l|l|}
\hline
    & $\sigma$    
    & $\pi_\perp$     
    & Internal symmetry & Translation\\ \hline
$\chi\gr{SB}$ &
    $\sigma\ne0$
    & $\pi_\perp=0$
    & $\gr{U}(1)_\gr{B}\times\gr{U}(1)_{\gr{Q}}$ 
    & Unbroken\\ \hline
PC 
    & $\sigma\ne 0$
    & $\pi\ne 0$     
    & $\gr{U}(1)_\gr{B}$ 
    & Unbroken \\ \hline
CDL   & $\sigma(z)$
    & $\pi_\perp=0$            
    & $\gr{U}(1)_{\gr{B}}\times\gr{U}(1)_{\gr{Q}}$
    & Broken \\ \hline
SPC   & $\sigma\ne0$ 
    & $\pi(z)$  
    & $\gr{U}(1)_{\gr{B}}$
    & Broken \\ \hline
\end{tabular}
}
\label{tab:states}
 \end{table}
%---------------------------TABLE---------------------------%

In Fig.~\ref{fig:phase_final} we display the GL
phase diagrams computed for $\muI^2=0.01$ (a), $0.1$ (b), $0.2$ (c), and
$0.5$ (d).
Let us start with the case $\muI^2=0.01$.
First, a rough order estimate is useful to have in mind what is
the physical scale of $\muI$.
Since $h\sim m_q/\Lambda$ with $m_q\sim10\,\gr{MeV}$ and
$\Lambda\sim1\,\gr{GeV}$
being the current quark mass and the energy scale for chiral symmetry
breaking, $\muI^2=0.01$ corresponds to $\muI=0.1\,\Lambda
(m_q/\Lambda)^{1/5}\sim40\,\gr{MeV}$.
We see that the LCP found in the previous analysis is intact apart
from the trivial shift of its location
$(\alpha_2^{\gr{LCP}},\alpha_4^{\gr{LCP}})%
\to(\alpha_2^{\gr{LCP}},\alpha_4^{\gr{LCP}}-\muI^2)$, which is
absorbed in the redefinition of the vertical axis:
$\alpha_4\to\alpha_4+\muI^2$.
The major topological change from the case $\muI=0$ is the appearance of
an island for the SPC replacing a part of the CDL phase which would have
extended off the LCP.
In fact, the second-order transition from the
$\sigma_S$-phase to the CDL is taken over by the one to the SPC for
$\alpha_4\alt-6.16$ where the instability for
developing an infinitesimal sinusoidal density wave of the charged pion
condensate takes place earlier than that for rippling the chiral
condensate.
This is because $\muI^2$ makes the coefficient of negative
gradient term $(\pi_\perp^\prime)^2$ larger than that of
$(\sigma^\prime)^2$ by ${\muI^2\alpha_6}/{3}$ as seen in
(\ref{eq:omegamuI}).
On the other hand, the SPC and CDL phases are separated by a first-order
transition.
As a consequence, there is a bicritical point marked by ``BCP'' where 
a first-order transition meets two second-order ones.

Let us briefly discuss what can be a possible interpretation of the
physical reason why an inhomogeneous pion condensate occurs at large
value of $\alpha_2$ which roughly corresponds to the high density side
of the $(\mu,T)$-phase diagram \cite{Abuki:2013vwa}.
The pion condensate for $\muI>0$ is described by the formation of
$u$ and $\bar{d}$ quark pair on the matched Fermi surface $\muI$
\cite{Son:2000xc}.
The effect of $\mu$ is to break the pair making mismatched Fermi surface
via producing a net excess of $u$ quarks over $\bar{d}$ quarks.
When this effect stresses the pair condensate, it could sometimes happen
that the pairing is broken partially within the real or momentum space 
such as in the FFLO superconductor in the presence of an external
magnetic field \cite{FFLO}. 

When $\muI^2$ increases to $0.1$, the situation changes to the one
displayed in Fig.~\ref{fig:phase_final}(b).
The CDL region shrinks and the SPC now occupies a major part.
The transition between the SPC and $\chi\gr{SB}$ phases is first-order,
accompanied by an abrupt change in $\sigma$.
Accordingly there shows up point ``E'' at which a second-order
transition comes across two first-order transitions.
Another notable change is the appearance of continent of PC
in the deep inside the $\chi$SB phase \cite{Abuki:2013vwa}; the two
phases are separated by a second-order transition.
Fig.~\ref{fig:phase_final}(c) shows the situation for $\muI^2=0.2$.
The PC now meets the SPC island, and their competition gives rise to a
first-order phase boundary between them.
As a result there appears a new Lifshitz point ``LP'', which
has two branches of first-order transitions, and a second-order
transition between the PC and $\chi\gr{SB}$ phases.
The phase diagram for $\muI^2=0.5$ is shown in
Fig.~\ref{fig:phase_final}(d).
The CDL region shrinks so much that its existence can
be only confirmed in the inset figure that magnifies the vicinity of 
the LCP.
The transition between the PC and $\chi\gr{SB}$ phases changes to
first-order before coming across the SPC island.
As a result, the LP now has three branches of first-order transitions.

\emph{Conclusion.--}
We investigated systematically the two-flavor QCD phase diagram near
the CP using a generalized version of GL approach, combined with the
perturbative expansion in $\muI$.
We have clarified that the effect of isospin
imbalance brings about drastic changes in the phase structure.
The most significant one is the stabilization of the SPC for a wide
range of GL parameter space.
Our results suggest that at low temperature, going down in density
from high density side, one may have a second-order phase transition
from the nearly chirally symmetric matter to the SPC, which is signaled by
development of a ripple sinusoidal density wave of a charged pion
condensate.
The state eventually evolves to solitonic lattice.

The magnetic property of inhomogeneous pion condensate is worth to be
addressed in the future.
The homogeneous PC has an electric charge, so is also a superconductor.
Thus it should exhibit a Meissner effect by which a weak magnetic field
applied to the system is expelled from the bulk.
In the inhomogeneous SPC phase, however, there are domainwalls where the
pair is effectively broken so that the magnetic field can penetrate
there.
This may bring some phenomenological consequences to compact star
physics.

There remain a couple of interesting questions unsolved.
First, it would be interesting to explore the possibility of higher
dimensional lattice structures. 
In particular it may be possible that the three dimensional spherical
chiral defect is formed in advance of the two-dimensional wall-like
defect studied here.
Second, it is interesting to specify the low energy excitations on
the SPC/CDL and clarify the physical nature of them.
Lastly, it should be worth try to extend the current GL analyses to
three-flavor case where we have to take care a possible Kaon condensate
driven by the chemical potential for strangeness.
The future study along these directions would make it clear how these
exotic inhomogeneous states leave unique footprints in the
phenomenology of compact star physics.

\vspace*{1ex}
The author thanks M.~Ruggieri for several useful comments.
Numerical calculations were carried out on SR16000 at YITP in
Kyoto University.

\bibliographystyle{elsarticle-num}
\bibliography{<your-bib-database>}

\end{document}